\documentstyle[preprint,aps,prl,epsfig]{revtex}

\newcounter{listfig}

\begin{document}


\title{Voltage-controlled tunneling anisotropic magneto-resistance of a ferromagnetic $p^{++}$-(Ga,Mn)As/$n^{+}$-GaAs
Zener-Esaki diode}

\author{R. Giraud, M. Gryglas, L. Thevenard, A. Lema\^itre, G. Faini}

\address{Laboratoire de Photonique et de Nanostructures - CNRS, Route de Nozay, 91460 Marcoussis, France}

\date{\today}

\maketitle

\begin{abstract}

The large tunneling anisotropic magneto-resistance of a single $p^{++}$-(Ga,Mn)As/$n^{+}$-GaAs Zener-Esaki diode is evidenced in
a perpendicular magnetic field over a large temperature and voltage range. Under an applied bias, the tunnel junction transparency 
is modified, allowing to continuously tune anisotropic transport properties 
between the tunneling and the ohmic regimes. 
Furthermore, an asymmetric bias-dependence of the anisotropic tunneling magneto-resistance is also observed:   
a reverse bias highlights the full (Ga,Mn)As valence band states contribution, whereas a forward bias 
only probes part of the density of states and reveals opposite contributions from two subbands.

\end{abstract}


\narrowtext

\vspace{1.5cm}

Diluted ferromagnetic semiconductors are considered as much promising candidates for future spin electronics
devices, as compared to ferromagnetic metals, showing versatile and specific functionalities \cite{Prinz98,Ohno99,Dietl00,Boukari02}, only limited by
still too low Curie temperatures to date. In particular, hole-induced ferromagnetism in a diluted magnetic
semiconductor, such as (Ga,Mn)As, is often associated with dominant anisotropic heavy-holes transport \cite{Abolfath01,Dietl01}. In an epitaxial thin film, 
distortions from the cubic zinc-blend structure, together with strong
spin--orbit couplings, lead to a highly anisotropic density of state at the
Fermi surface. Moreover, due to a large exchange interaction between delocalized holes and localized magnetic
Mn impurities, the Fermi states distribution evolves with the relative orientation of the magnetization with
respect to the crystallographic axis. In a magnetic tunnel junction, this gives rise to a new mechanism for tunneling magneto-resistance, the
so-called tunneling anisotropic magneto-resistance (TAMR), that adds up to the more familiar spin-conserving TMR
mechanism when the latter is also at work. 
Contrary to usual TMR, strain-induced tunneling AMR is almost independent of
spin flip events during the tunneling process, thus surviving even in presence of strong spin relaxation in the barrier.

Most important, novel spintronics tunneling devices solely based on a \emph{single} ferromagnetic layer can now be built, the
functionality of which fully relies on TAMR. Recently, a simple (Ga,Mn)As/AlO$_x$/Au tunnel junction indeed showed a significant
TAMR amplitude with a low-field spin-valve like signal of a few percent and of controllable sign \cite{Gould04}, a phenomenon unrevealed with 
ferromagnetic metals due to much smaller spin--orbit couplings. 
Roughly speaking, the tunnel junction acts as a filter in $k$-space for carriers that contribute to electrical transport,
thus limiting tunneling states to only a small fixed area of the Fermi surface around the tunneling direction.
Upon magnetization rotation, the tunneling density of states is modified by the Fermi surface states
redistribution, leading to a change in the tunnel resistance, that is TAMR. For a thin film with an easy plane
anisotropy, the magnetization can be continously and reversibly rotated by applying a magnetic field
perpendicular to the plane, whereas the magnetization orientation depends on both the in-plane anisotropy and
the hysteretic magnetization reversal processes when the field is applied within the plane. 
In both cases, a change in the device resistance is observed when the saturated magnetization lies 
either along or perpendicular to an anisotropy hard axis. 
Very recently, much larger
magneto-resistance (MR) amplitudes were also observed in nanostructures with two (Ga,Mn)As electrodes,
but in which other MR mechanisms coexist together with TAMR, both in vertical or lateral transport configurations \cite{Ruester05,Giddings05}.

In this letter, we describe an ultra-low power spintonics device, built on a single ferromagnetic $p^{++}$-(Ga,Mn)As/$n^{+}$-GaAs
Zener-Esaki tunneling junction, showing a large perpendicular-to-plane TAMR response, up to 40$\%$ at
$T=$~4.2~K. Contrary to previous reports, the TAMR signal occurs over a large temperature range, up to $T_c\approx$70~K, and a much wider voltage
bias range. This is respectively due to both a significant contribution from shape anisotropy (magnetization) in
the perpendicular configuration, and a large lever arm ratio as achieved with a degenerate $n^{+}$-GaAs
buffer layer lower doped than the $p^{++}$-(Ga,Mn)As one. Under reverse bias, the voltage-controlled tunneling barrier gives a convenient way to tune the device from the
tunneling to the ohmic regime, that is from a pure TAMR regime to the usual ohmic AMR one. This gives the first direct evidence of a continuous 
transition from AMR to TAMR when the interface resistance is increased, i.e. when the transparency of
the tunnel junction is reduced at lower bias. Furthermore, a strong asymmetry is observed in the bias dependence 
of the magneto-resistance, which is mainly related to the tunneling transport asymmetry (barrier, valence subbands density of states). 
Finally, very high accuracy dc-transport measurements, well-adapted to the 
tunneling regime, allow us to drive this ferromagnetic Zener-Esaki diode in the very low-bias regime, 
where TAMR is maximum and power consumption minimum.

The spin Zener-Esaki diode was grown in a molecular beam epitaxy chamber dedicated to (Ga,Mn)As-based heterostructures. It consists of a 
50~nm thick Ga$_{1-x}$Mn$_x$As epilayer, with $x\approx$~6\%, grown at low temperature ($T\approx$~250~$^{\circ}$C) over a Si-doped GaAs buffer layer 
($n^{+}\approx 2.10^{18}$~cm$^{-3}$) previously elaborated at high temperature ($T\approx$~500~$^{\circ}$C) onto an $n^{+}$-doped GaAs substrate of 
similar doping. Large square pillars ($\sim$400~nm high) were patterned by classic UV-lithography and wet etching techniques, with a surface ranging from 
(50~$\mu$m)$^2$ up to (300~$\mu$m)$^2$. All the measurements shown in this letter were obtained on the largest pillars, but perfect scaling 
was observed for these large dimensions, and transport properties were dominated by the interface resistance of the $p-n$ junction. Ohmic top and bottom contacts were respectively made by Ti/Au deposition on (Ga,Mn)As, and by In/Cu evaporation on $n$-GaAs 
followed by a low-$T$ annealing (2' at 200~$^{\circ}$C). 
From Hall bar measurements, the (Ga,Mn)As resistivity at $T=$~4.2~K is about 10m$\Omega$.cm and the Curie temperature is about 70~K. 
Ultra low-noise transport measurements were performed using a Keithley 6430 sub-femtoammeter SMU set with guarded two-probe coaxial wiring. 
A magnetic field up to 6000~Oe was produced by an electro-magnet. 

Fig1a) shows current-voltage characteristics of the Zener-Esaki diode at high (room) and low (liquid Helium) temperatures. The low-bias tunneling contribution 
dominates below the threshold voltages of about -1~V and +0.7~V, under reverse and forward bias respectively. 
Moreover, this ferromagnetic diode exhibits a clear magneto-current contribution at low temperature, as shown in 
Fig1b) in the tunneling regime under a constant reverse bias $V_{bias}=$-500~mV and for two perpendicular configurations of 
the saturated magnetization (in-plane, $H_{\perp}=0$~Oe; perpendicular-to-plane, $H_{\perp}^{sat}=6000$~Oe). 
Indeed, under a varying magnetic field applied perpendicularly to the tunnel interface ($\vec B \parallel \vec j$) 
and for $V_{bias}=$-500~mV, the device resistance continuously decreases from $R\sim$1.52~M$\Omega$ in zero-field (in-plane magnetization) down 
to $R\sim$1.34~M$\Omega$ at $H_{\perp}^{sat}=\pm$6000~Oe (perpendicular magnetization), thus giving a negative magneto-resistance of about 12\%. 
The TAMR response, defined as $[R(H_{\perp})-R(0)]/R(0)$, is shown in Fig1c) and the full range is obtained at $H_{\perp}=H_{\perp}^{sat}$. 
The signal evolves similarly to the perpendicular-to-plane component of the magnetization, being almost 
satured close to the anisotropy induction $B_A\approx$~4000~G, which is the sum of the saturated magnetization $M_S\sim$~1000~G (shape anisotropy) and the strain-induced perpendicular anisotropy $B^{\perp}_2\sim 3000$~G. 
This reversible behavior is typical of a magnetic thin film with an easy-plane anisotropy, and in which a perpendicular-to-plane anisotropy strenghtens 
demagnetizing field effects (as occurs in (Ga,Mn)As epilayers grown on GaAs \cite{Moore03}). 
Larger TAMR amplitudes were also observed at lower reverse bias, with an increasing amplitude up to 40~\% at $T=$~4.2~K and for $V_{bias}=$-1~mV. 
Such a small bias gives a huge junction resistance of about $10^{14}\Omega.\mu$m$^2$, 
which results in an ultra-low power consumption of about 1~fW. 

The full TAMR amplitude, $[R(H_{\perp}^{sat})-R(0)]/R(0)$, decreases almost linearly with increasing temperature, 
much as the magnetization itself, up to $T_c\approx$70~K where it vanishes, as observed for $V_{bias}=$-500~mV in Fig.~\ref{fig2}. 
A direct comparison is made with the (Ga,Mn)As planar Hall resistance \cite{Tang03} shown in the inset, 
which also gives a similar value of $T_c$. This clearly shows that perpendicular-to-plane TAMR directly 
scales with the magnetization projection along the growth direction, that is, along the anisotropy hard axis. Yet, 
a small contribution from the strain-induced perpendicular anisotropy to the temperature dependence of TAMR could also be at work.

As evidenced in Fig. 3, \emph{under reverse bias}, i.e. when spin-polarized holes from (Ga,Mn)As valence band tunnel to the $n$-GaAs conduction band, 
the TAMR amplitude decreases with increasing bias. This is in agreement with previous descriptions of a reduction of TAMR as the bias exceeds the exchange energy, making the dependence of the tunnel resistance 
on the magnetization orientation much smaller. Because the non-magnetic part of the diode has a much lower carrier density than the ferromagnetic one (roughly, two orders of magnitude), the potential mostly drops over the 
$n^{+}$-GaAs buffer, thus allowing us to probe the TAMR reponse up to a large bias of about $\pm$1~V (instead of about $\pm$10~mV in previous reports). 
This very strong lever arm ratio, of about 100, gives a very convenient way to probe the TAMR mechanism with a great voltage accuracy. 
In particular, the measurement clearly show two regimes in the reverse bias-dependence of TAMR, with a fast decrease at very small bias (up to $V_{bias}=$-10~mV) 
followed by a slower linear evolution. This crossover still remains unclear. Besides, at large applied voltages, the MR signal converges to a constant value 
which corresponds to the usual AMR response, as associated with the ohmic regime. Therefore, the bias is an efficient mean 
to tune the device from tunneling to ohmic transport, and Fig.3 gives a direct evidence of the \emph{continuous evolution from interfacial TAMR to bulk AMR}, 
with a clear transition around $V_{bias}=$-1~V.  

Strikingly, contrary to first observations \cite{Gould04,Ruester05,Giddings05}, the TAMR signal shows a \emph{non-monotonous behavior 
under forward bias}, and even changes its sign under a large-enough voltage polarization, well below the threshold voltage. 
Indeed, this is a direct consequence of inter-band tunneling transport. 
Firstly, the tunnel barrier is asymmetric, so that to get a current of comparable amplitude requires a larger forward than reverse bias, 
thus leading to a reduction of the TAMR amplitude under forward polarization (again due to a larger tunneling window). 
Secondly, and more important, only \emph{a small part} of the (Ga,Mn)As valence band states contributes to the tunneling current 
in forward bias, the tunneling window being limited by the Fermi energy in the $n^+$-GaAs conduction band 
(relative to the bottom of the conduction band). This is not the case under reverse bias, where every states matching the voltage bias window contribute to the tunneling current
(up to the Fermi energy, relative to the top of the valence band, that is, about 150~meV). 
Therefore, under forward bias, the Zener-Esaki diode allows us to do the \emph{spectroscopy of the TAMR mechanism}, 
continuously probing every subbands contributions from the Fermi level to the bottom of the valence band. This gives a unique opportunity to study the relative contribution of mixed heavy-light hole 
states to the TAMR process, which will be discussed elsewhere in detail. Note that, upon the magnetization rotation, the density of states redistribution is opposite for the third and the fourth subbands \cite{Gould04}, resulting in 
a change of sign of TAMR at low bias. Moreover, its amplitude decreases to zero at larger bias, due to a vanishing contribution of the first two subbands to anisotropic processes.

In summary, we have investigated the anisotropic magneto-transport properties of a ferromagnetic 
$p^{++}$-(Ga,Mn)As/$n^{+}$-GaAs Zener-Esaki diode by continously varying the interface resistance 
from the tunneling to the ohmic regimes. This device showed a large perpendicular tunneling anisotropic magneto-resistance 
which strongly depends on the details of inter-band tunneling. A very large lever arm ratio allows us to study the 
reverse bias-dependence of TAMR with a high accuracy. Under a forward bias, various 
(Ga,Mn)As valence subbands give opposite contributions to TAMR.

O.~Mauguin and L.~Largeau are aknowledged for X-ray characterization of the (Ga,Mn)As epilayers, 
and L.~Leroy for her technical assistance.

\newpage

\centerline{LIST OF FIGURES CAPTIONS}

\begin{list}{FIG.~\arabic{listfig}}{\usecounter{listfig}}

\item (a) Current-voltage characteritics of the diode at $T=$~300~K (dotted line) and $T=$~4.2~K (full line). 
The inset enhances the $I$($V$) curves at low-bias and $T=$~4.2~K. (b) Temperature dependence of the tunneling current under a constant reverse bias $V_{bias}=$-500~mV, in zero field (full line) and for 
a saturation field $H_{\perp}^{sat}=6000$~Oe perpendicular to the plane (dotted line). (c) Magneto-resistance at $T=$~4.2~K under $V_{bias}=$-500~mV 
as a function of the applied field $H_{\perp}$.

\item Temperature dependence of the perpendicular-to-plane anisotropic tunneling magneto-resistance under a constant reverse bias $V_{bias}=$-500~mV, as derived from Fig1b). 
Inset: temperature dependence of the (Ga,Mn)As planar Hall resistance, obtained from Hall bar measurements.

\item Asymmetric voltage bias dependence of the perpendicular-to-plane anisotropic tunneling magneto-resistance, as derived from measurements similar to Fig1a) (line) and Fig1c) (open squares).

\end{list}

\newpage

\begin{figure}
\centerline{\epsfxsize= 15 cm \epsfbox{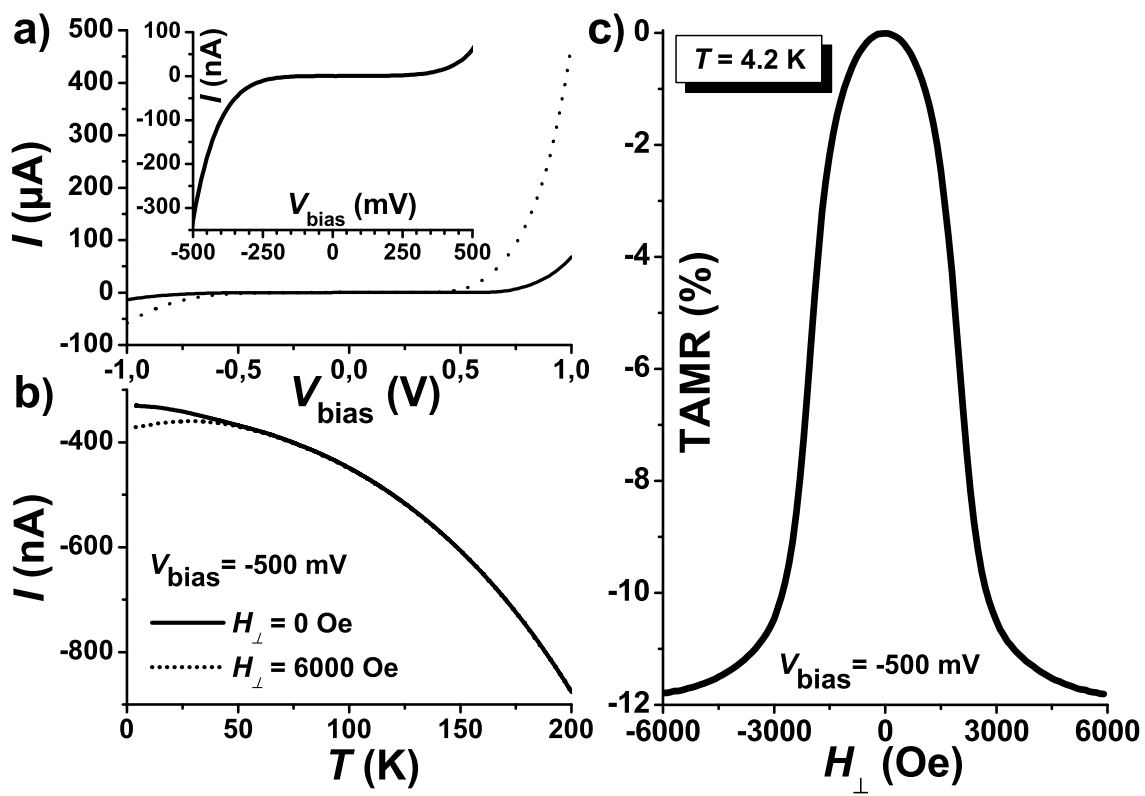}}
\caption{R. Giraud}
\label{fig1}
\end{figure}

\newpage

\begin{figure}
\centerline{\epsfxsize= 15 cm \epsfbox{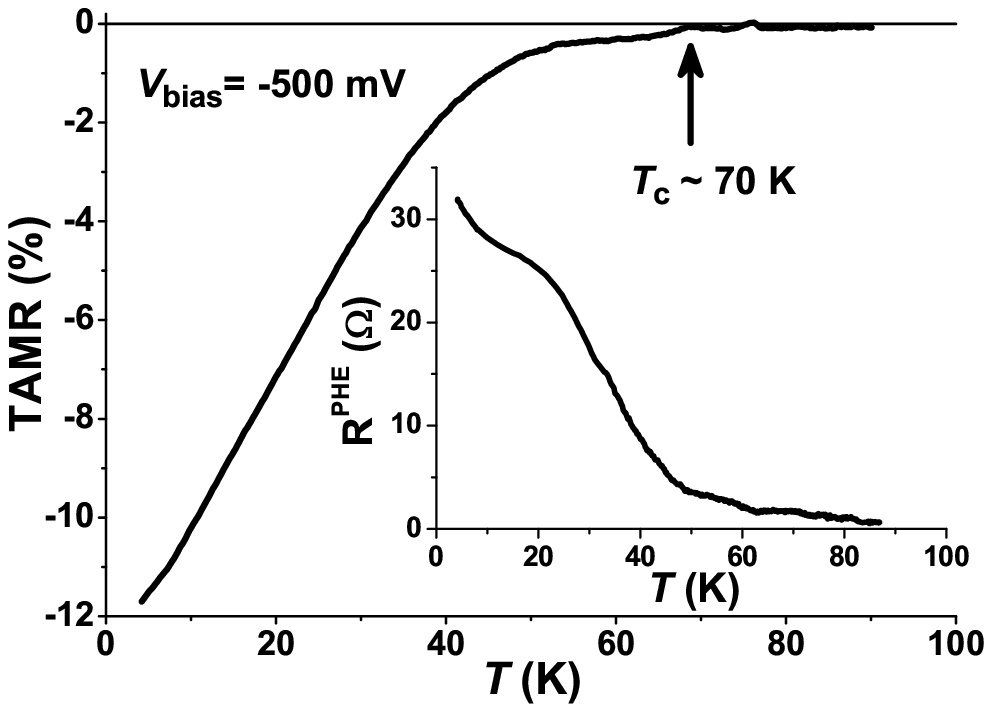}}
\caption{R. Giraud}
\label{fig2}
\end{figure}

\newpage

\begin{figure}
\centerline{\epsfxsize= 15 cm \epsfbox{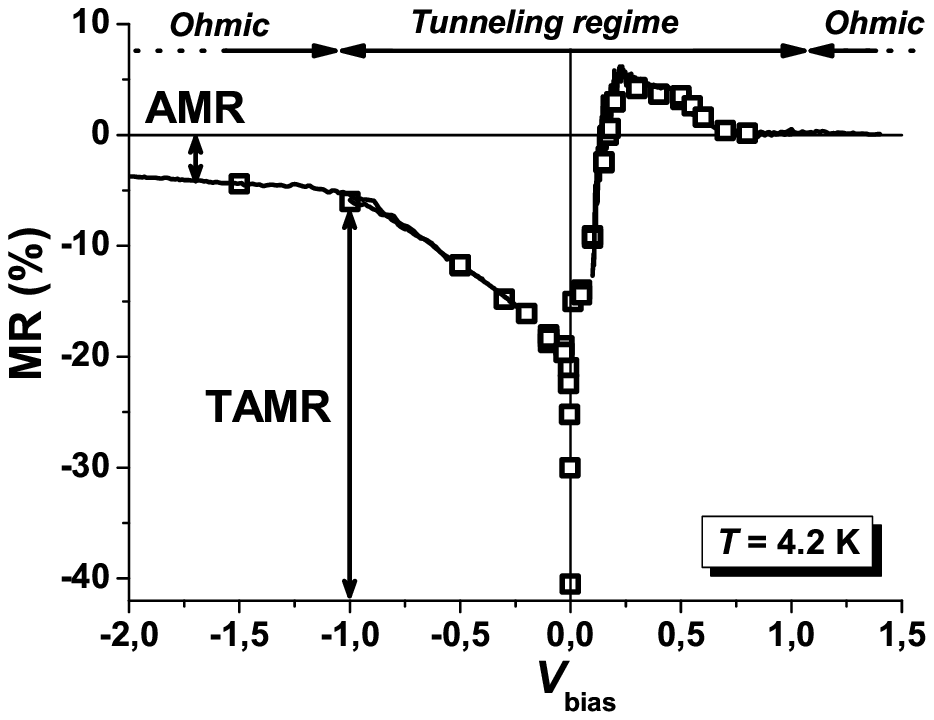}}
\caption{R. Giraud}
\label{fig3}
\end{figure}


\begin{references}

\bibitem{Prinz98}

G.~Prinz, Science \textbf{282}, 1660 (1998).

\bibitem{Ohno99}

H.~Ohno, J. Magn. Magn. Mater. \textbf{200}, 110 (1999).

\bibitem{Dietl00}

T.~Dietl {\it et al.}, Science \textbf{287}, 1019 (2000).

\bibitem{Boukari02}

H.~Boukari {\it et al.}, Phys. Rev. Lett. \textbf{88}, 207204 (2002).

\bibitem{Abolfath01}

M.~Abolfath {\it et al.}, Phys. Rev. B \textbf{63}, 054418 (2001).

\bibitem{Dietl01}

T.~Dietl, H.~Ohno, and F.~Matsukura, Phys. Rev. B \textbf{63}, 195205 (2001).

\bibitem{Gould04}

C.~Gould {\it et al.}, Phys. Rev. Lett. \textbf{93}, 117203 (2004).

\bibitem{Ruester05}

C.~Ruester {\it et al.}, Phys. Rev. Lett. \textbf{94}, 027203 (2005).

\bibitem{Giddings05}

A.D.~Giddings {\it et al.}, Phys. Rev. Lett. \textbf{94}, 127202 (2005).

\bibitem{Moore03}

G.P.~Moore {\it et al.}, J. Appl. Phys. \textbf{94}, 4530 (2003).

\bibitem{Tang03}

H.X.~Tang {\it et al.}, Phys. Rev. Lett. \textbf{90}, 107201 (2003).

\end{references}
\end{document}